\documentclass{ws-ijmpd}

\begin{document}

\markboth{Sergio del Campo and Jos\'{e} Villanueva}
{Observational constraints on the Generalized Chaplygin Gas}

\title{OBSERVATIONAL CONSTRAINTS ON THE GENERALIZED CHAPLYGIN GAS}

\author{ SERGIO DEL CAMPO }

\address{Instituto de F\'{\i}sica, Pontificia Universidad de Cat\'{o}lica de
Valpara\'{\i}so, Casilla 4950\\
Valpara\'{\i}so , Chile\\
sdelcamp@ucv.cl}

\author{ J.R.VILLANUEVA }

\address{Instituto de F\'{\i}sica, Pontificia Universidad de Cat\'{o}lica de
Valpara\'{\i}so, Casilla 4950\\
Valpara\'{\i}so , Chile\\
jose.villanueva.l@mail.ucv.cl}

\maketitle

%\begin{history}
%\received{Day Month Year}
%\revised{Day Month Year}
%\comby{Managing Editor}
%\end{history}

\begin{abstract}
In this paper we study a quintessence cosmological model in which
the dark energy component is considered to be the Generalized
Chaplygin Gas and the curvature of the three-geometry is taken
into account. Two parameters characterize this sort of fluid, the
$\nu$ and the $\alpha$ parameters. We use different astronomical
data for restricting these parameters. It is shown that the
constraint $\nu \lesssim \alpha$ agrees enough well with the
astronomical observations.
\end{abstract}


\section{Introduction}

Current measurements of redshift and luminosity-distance relations
of Type Ia Supernovae (SNe) indicate that the expansion of the
Universe presents an accelerated phase \cite{R98,P99}. In fact,
the astronomical measurements showed that Type Ia SNe at a
redshift of $z \sim 0.5$ were systematically fainted which could
be attributed to an acceleration of the universe caused by a
non-zero vacuum energy density. This  gives as a result that the
pressure and the energy density of the universe should violate the
strong energy condition, $\rho_X + 3\,p_X\,>\,0$, where $\rho_X$
and $p_X$ are energy density and pressure of some matter
denominated dark energy, respectively. A direct consequence of
this, it is that the pressure must be negative. However, although
fundamental for our understanding of the evolution of the
universe, its nature remains a completely open question nowadays.

Various models of dark energy have been proposed so far. Perhaps,
the most traditional candidate to be considered is a non-vanishing
cosmological constant \cite{S03,T04}. Other possibilities are
quintessence \cite{C98,Z98}, k-essence \cite{Ch00,AP00,AP01},
phantom field \cite{C02,C03,H04}, holographic dark energy
\cite{Z07,W07}, etc. (see ref.~\refcite{D09} for model-independent
description of the properties of the dark energy and ref.~\refcite{S09} for possible alternatives).

One of the possible candidate for dark energy that would like to
consider here is the so-called Chaplygin gas (CG) \cite{b1}. This
is a fluid described by a quite unusual equation of state, whose
characteristic is that it behaves as a pressureless fluid at the
early stages of the evolution of the universe and as a
cosmological constant at late times. Actually, in ref.~\refcite{K01},
it was recognized its relevance to the detected cosmic
acceleration. They found that the CG model exhibits excellent
agreement with observations. From this time, the cosmological
implications of the CG model have been intensively investigated in
the literature \cite{De03,G03,A05,Z05}. Subsequently, it was
notice that this model can be generalized, which now it is called
the generalize Chaplygin gas (GCG). This GCG model was introduced
in ref.~\refcite{K01} and elaborated in ref.~\refcite{B02}. After these works,
the cosmological implications of the GCG model have been
intensively investigated in the literature \cite{A03,M03,P04,Z04,B04,B05,Ch05,G05,F08}.
There are claims that it does not pass the test connected with structure formation because of predicted
but not observed strong oscillations of the matter power spectrum \cite{STZW04}. It should be mentioned, however, the
oscillations in the Chaplygin gas component do not necessarily imply corresponding oscillations in the
observed baryonic power spectrum \cite{BACM03}. This is a topic that requires much more studies.
It is was realized that these kind of models have a clearly stated connection with
high-dimension theories \cite{F02}. Here, the GCG appears as an
effective fluid associated with d-branes. Also, at the fundamental
level, it could be derived from the Born-Infeld action
\cite{B03}.

On the other hand, today we do not know precisely the geometry of
the universe, since we do not know the exact amount of matter
present in the Universe. Various tests of cosmological models,
including space–time geometry, galaxy peculiar velocities,
structure formation and very early universe descriptions (related
to the Guth´s inflationary universe model \cite{G81}) support a
flat universe scenario. Specifically, by using the five-year
Wilkinson Microwave Anisotropy Probe (WMAP) data combined with
measurements of Type Ia supernovae (SN) and Baryon Acoustic
Oscillations (BAO) in the galaxy distribution, was reported the
following value for the total matter density parameter,
$\Omega_T$, at the 68\% CL uncertainties, $\Omega_T = 1.02 \pm
0.02$ \cite{H09}.

In this respects we wish to study universe models that have
curvature and are composed by two matter components. One of these
components is the usual nonrelativistic dark  matter (dust); the
other component corresponds to dark energy which is supposed to be
a sort of quintessence-type matter, described by a Chaplygin
gas-type, or more specifically the GCG.

We should mention that in what concern with the Bayesian analysis
the cosmological constant is favored over GCG \cite{SKK06,KS08,KS07}. However, in ref. \cite{LWY09},
it was shown that the GCG models, proposed as candidates of the unified dark matter-dark energy (UDME),
are tested with the look-back time (LT) redshift data. They found that the LT data only give a very weak
constraint on the parameters. But, when they combine the LT redshift data with the baryonic acoustic
oscillation peak the GCG appears as a viable candidate for dark energy. On the other hand, the GCG model
has been constrained with the integrated Sach-Wolf effect. Recently, a gauge-invariant analysis of the baryonic
matter power spectrum for GCG cosmologies was shown to be compatible with the data \cite{GKMPS08,F08,FGVZ08}.
This result seems to strengthen the role of Chaplygin gas type models as competitive candidates for the dark sector.

Our paper is organized as follow: In section II we present the
main characteristic properties and we introduce some definition
related to the GCG. In section III we study the kinematics of our
model. Here, we take quantities such that the modulus distance,
luminosity distance, angular size, among others. In section IV we
proceed to describe the so-called shift parameter which is related
to the position of the first acoustic peak in the power spectrum
of the temperature anisotropies of the cosmic microwave background
(CMB) anisotropies. We give our conclusions in Section V.

\section{The Generalized Chaplygin Gas (GCG)}

Let us star by considering the equation of state (EOS)
corresponding to the GCG

\begin{equation}
p_{gcg} = - \nu  \frac{\Xi}{\rho_{gcg}^{\alpha}}\,. \label{gcg}
\end{equation}

Here, $p_{gcg}$ and  $\rho_{gcg}$ are the pressure and the energy
density related to the GCG, respectively. $\nu$ is the square of
the actual speed of sound in the GCG and $\alpha$ is the GCG
index. $\Xi$ is a function of $\alpha$ and $\rho_{gcg}^{(0)}$ (the
present value of the energy density of the GCG), and it is given
by

\begin{equation}
\Xi \equiv \Xi(\rho_{gcg}^{(0)} \mid \alpha)= \frac{1}{\alpha}
\left(\rho_{gcg}^{(0)}\right)^{1+\alpha}.\label{gcg0}
\end{equation}

The dimensionless energy density related to the GCG
$f_{gcg}(z;\nu,\alpha) \equiv \rho_{gcg}(z;\nu,\alpha) /
\rho_{gcg}^{(0)}$ becomes given as a function of the red shift,
$z$, and the parameters $\alpha$ and $\nu$ as follows

\begin{equation}
f_{gcg}(z;\nu,\alpha)= \left[\frac{\nu}{\alpha} +
\left(1-\frac{\nu}{\alpha}\right)(1+z)^{3(1+\alpha)}\right]^{\frac{1}{1+\alpha}}.
 \label{gcg1}
\end{equation}

Here, we have considering a Friedmann-Robertson-Walker (FRW)
metric, and we have used the energy conservation equation:
${\frac{d\rho_{gcg}}{dt}+3H\left(\rho_{gcg}+p_{gcg}\right) = 0}$,
where $H$ represents the Hubble factor. Note that if $\alpha =
\nu$, we get that $f_{gcg}(z,\alpha)=1$ (the same happen for
$z=0$), which means that the energy density related to the GCG
corresponds to a cosmological constant.

\begin{figure}[pb]
\centerline{\psfig{file=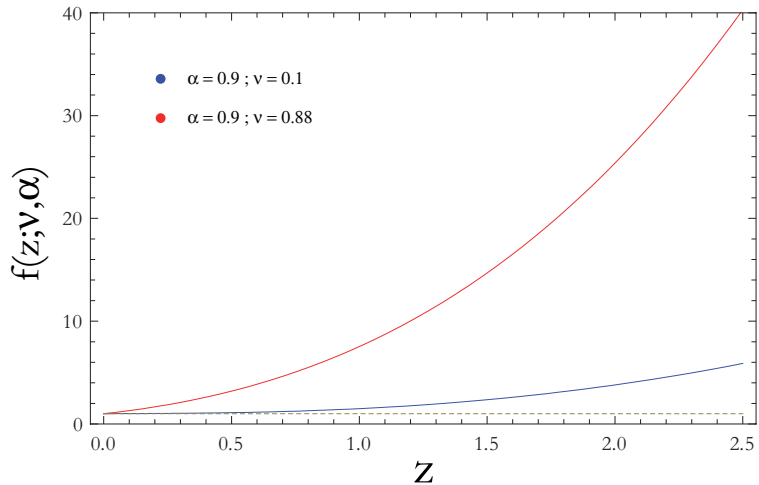,width=10cm}}
  \vspace*{8pt}
\caption[]{Plot of the function $f_{gcg}(z;\nu,\alpha)$ as a
function of the red shift, $z$ for the cases $\alpha > \nu $ (
$\alpha = 0.9$ ; $\nu = 0.1 $, blue line) and $\alpha \approx \nu
$ ($\alpha = 0.9$; $\nu = 0.88 $, red line). These two cases are
compared with that corresponding to the cosmological constant,
$\Lambda $, case (dashed line).} \label{Fig01}
\end{figure}

In Fig.\ref{Fig01} we plot $f_{gcg}(z;\nu,\alpha)$ as a function
of the red shift, $z$. Note that this function is highly sensitive
to the difference between the values of $\alpha$ and $\nu$.

The derivative of the function $f_{gcg}(z;\nu,\alpha)$ with
respect to the redshift, $z$, becomes given by
%\begin{equation}
$$
\frac{d f_{gcg}(z;\nu,\alpha)}{dz}\equiv f'_{gcg}(z;\nu,\alpha)=
3\left(1-\frac{\nu}{\alpha}\right) \frac{(1+z)^{3\alpha+2}}
{f^{\alpha}(z;\nu,\alpha)}\,. \label{gcg2}
$$
%\end{equation}
Note that the sign of this function depends on the values that the
constants $\nu$ and $\alpha$ could take . For $\nu \gtrless
\alpha$ we have that $f'_{gcg}(z;\nu,\alpha) \lessgtr 0$. We need
to have that the function $f_{gcg}(z;\nu,\alpha)$ to be greater
that zero, since it corresponds to an energy density in an
expanding universe. Thus, we expect that the case $\nu < \alpha$
be relevant for our study.

We can write the EOS related to the GCG in the barotropic form as
follows
\begin{equation}
p_{gcg}= \omega_{gcg} \rho_{gcg}, \label{gcg3}
\end{equation}
where the equation of state parameter, $\omega_{gcg}(z)$, becomes
given by
\begin{equation}
\omega_{gcg}(z) = - \frac{\nu / \alpha}{\frac{\nu}{\alpha} +
\left(1-\frac{\nu}{\alpha}\right)(1+z)^{3(1+\alpha)}}.
\label{gcg4}
\end{equation}

\begin{figure}[pb]
\centerline{\psfig{file=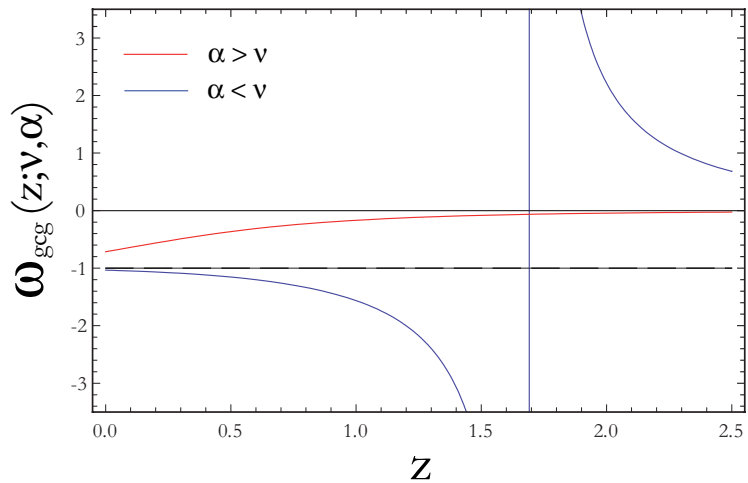,width=10cm}}
  \vspace*{8pt}
  \caption{Plot of the EOS parameter, $\omega(z;\nu,\alpha)$, as a function of the red shift,
  $z$.  This function for $\nu<\beta$ lies in the range between $-\nu/\beta$ (for $z=0$) and
  $0$ (for $z\longrightarrow \infty$). For $\nu=\beta$ this parameter gets the value $-1$,
  and for $\nu>\beta$  this parameter present two branches (one positive and the other
  negative). It becomes $\omega_{gcg}\longrightarrow \mp \infty$
  at some specific value of the red shift, $z=z_c$.}
  \label{Fig02}
\end{figure}

Of course, for $\nu = \alpha=0$ we get  $\omega_{gcg} = -1$,
corresponding to the cosmological constant case. In Fig.\ref{Fig02}
we have plotted the EOS parameter, $\omega(z;\nu,\alpha)$, as a
function of the red shift, $z$. Note that for
$z_c=\left(\frac{1}{1-\alpha/\nu}\right)^{\frac{1}{3(1+\alpha)}}-1$,
with $\alpha < \nu$, the EOS parameter, $\omega(z_c;\nu,\alpha)$,
goes to  minus (plus) infinity, i.e
$\omega(z_c;\nu,\alpha)\longrightarrow \mp\infty$. The minus
(plus) sign corresponds to the $z<z_c$ ($z>z_c$) branch. These
situations are represented in Fig.\ref{Fig02} by the blue lines.
For an accelerating phase of the universe we need to take into
account the $z<z_c$ branch only, since it gives the right negative
sign for the EOS parameter. For $\alpha > \nu$, the EOS parameter
always is negative, i.e. $-\frac{\nu}{\alpha}\leq\omega_{gcg}<0$.
Summarizing, we can see from the latter equation that for $\nu >
\alpha$ we have $-1 < -\nu / \alpha \leq \omega_{gcg} < 0$ and for
$\nu < \alpha$ we find that $-1> - \nu / \alpha \geq \omega_{gcg}
> - \infty$.

A Taylor expansion of the EOS parameter, $\omega_{gcg}(z)$, around
$z=0$ becomes

\begin{equation}
 \omega_{gcg}(z)= -
\beta+3\beta(1-\beta)(1+\alpha)z -3\beta (1-\beta)
(1+\alpha)  \left[ 3(1-2
\beta)(1+\alpha)+1\right]z^2
 + O(z^3) , \label{expOmega}
\end{equation}

where $\beta=\frac{\nu}{\alpha}$.

In a spatially flat universe, the combination of WMAP and the
Supernova Legacy Survey (SNLS) data leads to a significant
constraint on the equation of state parameter for the dark energy
$ w(0) = -0.967^{+0.073}_{-0.072}$ \cite{S07}. This constraint
restricts the value of the ratio $\frac{\nu}{\alpha}$. The value
of this ratio used above, (see Fig. \ref{Fig01}), lies inside the
observational astronomical range of the parameter $\omega(0)$. The
case in which the EOS parameter is a linear function of the
redshift was studied in \cite{C99,D03}. This, it is a good
parametrization at a low redshift.

Phenomenological models of a specific time dependent
parametrization of the EOS, together with a constant speed of
sound have being described in the literature. A simple example is
the parametrization expressed by the EOS \cite{Ch01,L03} $
\omega(z) = \omega(0) + \frac{d\omega(z)}{dz}
\biggr{|}_{0}\frac{z}{(1+z)}$ corresponding to non-interacting
dark energy. By matching this parametrization with our expression
at low redshift we find that the parameter $\frac{d\omega(z)}{dz}
\biggr{|}_{0}$ and $3\beta(1-\beta)(1+\alpha)$ coincides. The
determination of the dynamical character of the EOS parameter,
$\omega(z)$, becomes important in future experiments. This
relevance has been notice by the The Dark Energy Task Force (DETF)
\cite{A06}. The coming decade will be an exciting period for dark
energy research.

\section{KINEMATICS OF THE MODEL}

In order to describe some important distances we introduce the
dimensionless Hubble function, $E(z)= \frac{H(z)}{H_0}$, reads as
\begin{equation}
 E^{2}(z;\nu,\alpha)  =
 \Omega_{cdm}^{(0)} (1+z)^{3} + \Omega_{k}^{(0)} (1+z)^{2}
  + \Omega_{gcg}^{(0)} f_{gcg}(z;\nu,\alpha) ,\label{gcg5}
\end{equation}

\noindent where $\Omega_{k}^{(0)} = -k/H_{0}^{2}$, and $\Omega_{cdm}^{(0)}$
and $\Omega_{k}^{(0)}$ represent the present cold dark matter and
curvature density parameters, respectively. Here. the parameter
$k$ takes the values $-1$, $0$ or $+1$, for open, flat or closed
geometries, respectively. $H_0 \equiv H(0) = 100 h \,km \,s^{-1}
Mpc^{-1}$ is the current value of the Hubble parameter. The
$E(z;\nu,\alpha)$ quantity depends on the values of the parameters
$\alpha$ and $\nu$, apart of the actual values of the density
parameters, $\Omega_{k}^{(0)}$, $\Omega_{cdm}^{(0)}$ and $
\Omega_{gcg}^{(0)}$. Note that these latter parameters satisfy the
constraint $\Omega_{k}^{(0)}+\Omega_{cdm}^{(0)}+
\Omega_{gcg}^{(0)} = 1$. On the other hand, astronomical
measurements will constraint the $\alpha$ and $\nu$ parameters, as
we will see.

\begin{figure}[pb]
  \centerline{\psfig{file=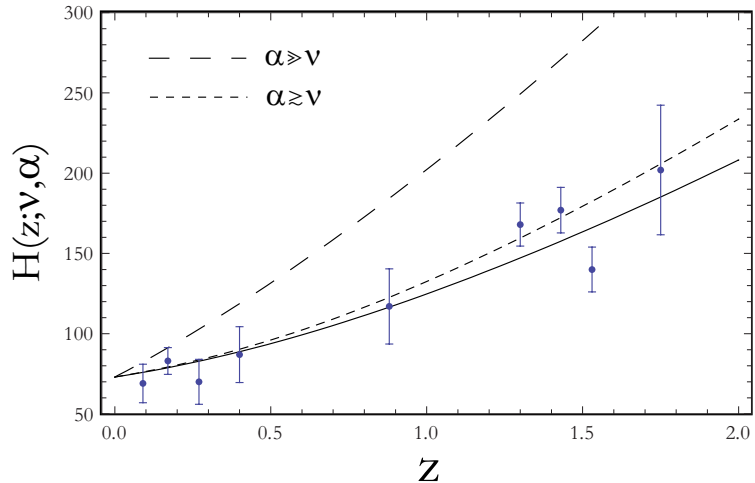,width=10cm}}
    \vspace*{10pt}
  \caption[]{Plot of the Hubble parameter, $H(z;\nu,\alpha)$, as a
  function of the redshift,  $z$. Here, we have introduced the
  observational values for the   Hubble's parameter (see ref.~\refcite{S09}) .
  The analytical curves were determined by using  $H_{0} = 73 [\textrm{Mpc}^{-1}
  \textrm{Km}/ \textrm{s}]$ for the present value of the Hubble's parameter
  and   we have taken $\Omega_{gcg}^{(0)}=0.725$ and $\Omega_{cdm}^{(0)}=0.275$
  for a flat geometry. The two GCG
  curves (small and large dashing) were plotted by taking $\alpha =0.9
  \gg \nu =0.1 $ and  $\alpha =0.9 \gtrsim \nu = 0.88$.
  The solid line represents the $\Lambda$CDM model.}
  \label{Fig03}
\end{figure}

In FIG.\ref{Fig03} we have taken $\Omega_{gcg}^{(0)}=0.725$ and
$\Omega_{cdm}^{(0)}=0.275$ with $\Omega_{k}^{(0)}=0$ for the
theoretical curves, and we have introduced the observational
values for the Hubble parameter from Ref.~\refcite{S06}.
The curves were plotted for both regimes, $\alpha \gtrsim
\nu $ ( $\alpha = 0.9$ and $\nu = 0.88$) and $\alpha \gg \nu $
($\alpha = 0.9$ and $\nu = 0.1$). In order to compare these curves
with the standard model we have included the $\Lambda$CDM model,
also. Note that the curve with $\alpha \gtrsim \nu $ is closed to
the observational data than that curve corresponding to $\alpha
\gg \nu$. Thus, when curvature is present into the cosmological
model, the curve with $\alpha \gtrsim \nu $ competes with the
standard cosmology (the $\Lambda$CDM model), in this respect. Note
also that no difference between the $\Lambda$CDM model and that
model were a GCG is included together with the curvature is found
for low redshift.

\subsection{Luminosity distance - redshift}

One of the more important observable magnitudes that we will
consider here will be luminous distance, $d _ {L} $. This is
defined as the ratio of the emitted energy per unit time,
$\mathcal{L}$, and the energy received per unit time $\mathcal{F}$
\cite{P03}

 \begin{equation}
d_L=\frac{\mathcal{L}}{4 \pi \mathcal{F}}. \label{dl01}
 \end{equation}
In this way, the luminosity distance can be written as

\begin{equation}
d_L(z; \nu, \alpha) = H_{0}^{-1} (1+z) y(z; \nu,
\alpha),\label{mo1}
\end{equation}

\noindent where the function $y(z; \nu, \alpha)$ becomes given by

\begin{equation} y(z; \nu, \alpha)= \frac{1}{\sqrt{\left|\Omega_{k}^{(0)}\right|}} \;
S_{k} \left\{ \sqrt{\left|\Omega_{k}^{(0)}\right|}  \int_0^z
\frac{d z'}{ E(z';\nu,\alpha)} \right\},\label{mo2}\end{equation}

\noindent and $S_k(x)$ takes the following expression for the
different values of the parameter $k$,

\begin{equation}
S_k(x) \left\{
\begin{array}{ll}
       \sin(x), & k=+1;\\
        x,       & k=0;\\
        \sinh(x), & k=-1.\\
\end{array} \right.\label{mo3}
\end{equation}

\begin{figure}[pb]
\centerline{\psfig{file=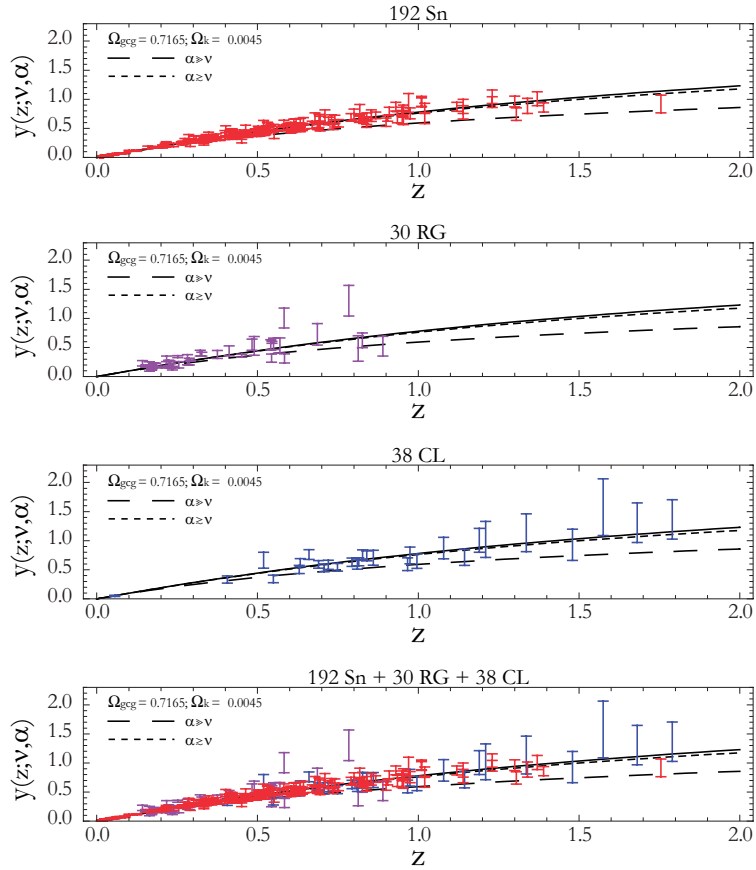,width=10cm}}
  \vspace*{15pt}
\caption{Plots of the theoretical curves for $y(z;\nu, \alpha)$
as a function of the redshift, $z$ for two different regimes:
$\nu=0.1 \ll \alpha =0.9$ and $\nu = 0.88 \lesssim \alpha =0.9$.
These curves are compared with astronomical data extracted from
Daly {\it et al} 2007; Left top: 192 Supernovas (Sn); Right top: 30
Radio Galaxies (RG); Left down: 38 Galaxy Clusters (CL); Right
down: 192 Sn + 30 RG + 38 CL. Here, we have taken the values
$\Omega_{k}^{(0)}=0.0045$, $\Omega_{gcg}^{(0)}=0.7165$ and
$\Omega_{cdm}^{(0)}=0.2790$.} \label{Fig04}
\end{figure}

By using the samples of $192$ supernova standard candles, $30$
radio galaxy and $38$ cluster standard rulers, presented in ref.~\refcite{D07},
we check our model described by Eq. (\ref{mo2}). This
check is done under the assumption that the curvature density
parameter, $\Omega_{k}^{(0)}$ takes the value
$\Omega_{k}^{(0)}=0.0045$ and the other parameters are
$\Omega_{gcg}^{(0)}=0.7165$ and $\Omega_{cdm}^{(0)}=0.2790$.
Fig.\ref{Fig04} shows some curves related to our model. It is
clear that the range of parameters for the GCG, as before, it is
near to the limit $\alpha \gtrsim \nu$, better that the limit
$\alpha \gg \nu$. Nevertheless, we cannot discriminate with
facility when we compare our curves (for the $\alpha \gtrsim \nu$
case) with that corresponding to the $\Lambda$CDM model. However,
this comparison becomes indistinguishable for small redshift, i.e.
$z \lesssim 0.7$.

One interesting quantity related to the luminosity distance,
$d_L$, is the distance modulus, $\mu$, which is defined as
\cite{V04}

\begin{equation}
\mu = {5} \; \log_{10}[d_L/(1 \hbox{Mpc})] +25.
\end{equation}

\begin{figure}[pb]
\centerline{\psfig{file=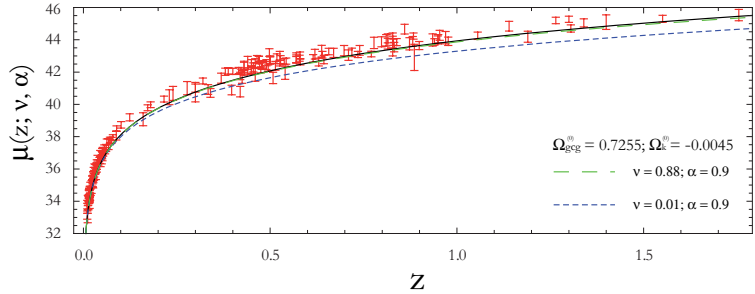,width=10cm}}
  \vspace*{10pt}
\caption{Graphic representing the magnitude $\mu(z;\nu,\alpha)$ as
a function of the redshifts, $z$. Here we have plotted two curves,
one for $\nu \lesssim \alpha$ ($\alpha=0.9$ and $\nu=0.88$) and
the other one for $\nu \ll \alpha$ ($\nu=0.01$ and $\alpha =
0.9$). Here, we have taken the values $\Omega_{gcg}^{(0)}=0.7255$
and $\Omega_{cdm}^{(0)}=0.279$. Also, we have included in this
plot the $\Lambda$CDM model, with $\Omega_{\Lambda}^{(0)} =
\Omega_{gcg}^{(0)}=0.7255$. The data were taken from Riess {\it et al} 2004.}
\label{fig:fig.mu}
\end{figure}

In Fig.\ref{fig:fig.mu} we have plotted $\mu$ as a function of the
redshifts, $z$. The values for the different GCG parameters are
the two set:  $\alpha=0.9$ and $\nu=0.88$, and $\alpha=0.9$ and
$\nu=0.01$. In each case we have considered that
$\Omega_{cdm}^{(0)}=0.279$, $\Omega_{gcg}^{(0)}=0.7255$ and
$\Omega_{k}^{(0)}=-0.0045$. Also, we have included in this plot
the $\Lambda$CDM model, with $\Omega_{\Lambda}^{(0)} =
\Omega_{gcg}^{(0)}=0.7255$. The data included in this graph were
taken from ref.~\refcite{R04}. Note that the case for $\nu \lesssim
\alpha$ becomes practically indistinguishable from that
corresponding to the $\Lambda$CDM model.

\subsection{Angular size - redshift}

The angular size, $\Theta$, is defined as the ratio of an object´s
physical transverse size, $l$, to the angular diameter distance
,$d_{A}$. This latter distance is related to the luminosity
distance, $d_L$ by mean of the relation $d_A= d_L/(1+z)^2$.
Therefore, we have

\begin{equation} \Theta(z;\nu,\alpha)\equiv\frac{l}{d_A(z;\nu,\alpha)} = \kappa
\frac{1+z}{y(z;\nu,\alpha)},\label{as}\end{equation}

\noindent Here, $l=l_{0}h^{-1}$, with $l_{0}$ the linear size
scaling factor and $\kappa=l H_{0}/c = 0.432 l_{0} [mas/pc]$.

Following our treatment of the comparison of the chaplygin gas
with the available data, we use the ref.~\refcite{G99} compilation
into 12 bins with 12-13 sources which satisfies the conditions in
which the spectral index lies in the range $-0.38 \leq \eta \leq
0.18$ and a total radio luminosity, $L$, which satisfies the
constraint, $L h^{2} \geq 10^{26} [W/Hz]$.

\begin{figure}[pb]
\centerline{\psfig{file=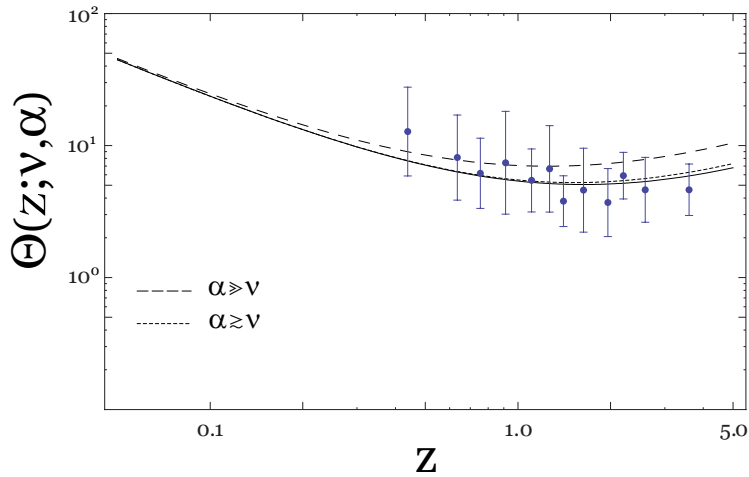,width=10cm}}
\vspace*{8pt}
  \caption[]{The angular size, $\Theta$, as a function of the redshift, $z$.
 The curves were determined by using   the value $l_{0}=4.86 [pc]$ and
$\Omega_{cdm}^{(0)}=0.2790$, $\Omega_{gcg}^{(0)}=0.7255$
$\Omega_{k}^{(0)}=-0.0045$. The data correspond to 145 sources
compiled by Gurvits {\it et al} 1999.}
 \label{Fig05}
\end{figure}

This points are showed in FIG.\ref{Fig05} together with the curves
determined by taking the values $l_{0}=4.86 [pc]$
$\Omega_{cdm}^{(0)}=0.2790$, $\Omega_{gcg}^{(0)}=0.7255$
$\Omega_{k}^{(0)}=-0.0045$. Note once again that the case for
which $\alpha \gtrsim \nu $ becomes favored than that the case
corresponding to $\alpha \gg \nu $. Here, as before we have
include the case corresponding to the $\Lambda$CDM model specified
by a continuous line.

\subsection{Deceleration, jerk, and snap parameters - redshift}

The luminosity distance, $d_L$, could be expanded in such a way
that the first Taylor coefficients of this expansion are related
to the parameters denominated deceleration ($q$), jerk ($j$), and
snap ($s$) parameters evaluated at present time. These three
parameters are defined in term of the second, third, and fourth
derivatives of the scale factor with respect to time,
respectively. The expansion of $d_L$ in term of the redshift, $z$,
reads\cite{V04}

\begin{eqnarray}
\displaystyle d_L(z) = & \hspace{-1cm}\frac{c
z}{H_0}\left\{1^{}_{} +\frac{1}{2}\left[1-q_0\right]z -
\frac{1}{6}\left[1-q_0-3q_0^2+j_0 \right. \right.\nonumber
\\&\hspace{-1cm}\left. +\frac{k c^2}{H_0^2 a_0^2}\right]z^2
+\frac{1}{24}\left[2-2q_0-15q_0^2-15\,q_0^3+5j_0+10 \,q_0 j_0
\right. \nonumber
\\ &   \left. \left. + s_0+\frac{k c^2(1+3q_0)}{H_0^2 a_0^2}\right]z^3
+ O(z^4) \right\}.
\end{eqnarray}

For our model the deceleration parameter, $q(z;\nu,\alpha)$
becomes given by

\begin{eqnarray}
q(z;\nu,\alpha) &=& -1 +
\frac{(1+z)E'(z;\nu,\alpha)}{E(z;\nu,\alpha)} \nonumber\\
&&=\frac{1}{2}\left[1 -
\frac{3\frac{\nu}{\alpha}\Omega_{gcg}^{(0)}f^{-\alpha}(z;\nu,\alpha)+
\Omega_{k}^{(0)}(1+z)^{2}}{E^{2}(z;\nu,\alpha)}\right]. \label{q}
\end{eqnarray}

\noindent The present value of this parameter becomes

\begin{equation}
q(0;\nu,\alpha)\equiv q_0(\nu,\alpha) =
\frac{1}{2}\left[\Omega_{cdm}^{(0)}-\left(3\frac{\nu}{\alpha}-1\right)\Omega_{gcg}^{(0)}\right].
\label{q0}\end{equation}
%Note that this parameter becomes independent of the
%curvature density parameter, $\Omega_k^{(0)}$.

In order to describe an accelerating universe, we need to satisfy
the constraint
$$
\frac{\nu}{\alpha} > \frac{1}{3}\left(1 +
\frac{\Omega_{cdm}^{(0)}}{\Omega_{gcg}^{(0)}}\right).
$$
Taking the ratio
$\frac{\Omega_{cdm}^{(0)}}{\Omega_{gcg}^{(0)}}\approx \frac{3}{7}$
we get that the $\nu$ and $\alpha$ parameters must satisfy the
bound $\frac{\nu}{\alpha}> \frac{10}{21}$. Note that the values of
this ratio that better agree with the astronomical data described
previously satisfy this restriction, since in most of them we have
taken $\nu = 0.88 \lesssim \alpha =0.9$.

With respect to the jerk, $j$, parameter we have that this becomes
given by

\begin{equation} j(z;\nu,\alpha)=3 q^{2}(z;\nu,\alpha)
+ \frac{(1+z)^{2}E''(z;\nu,\alpha)}{E(z;\nu,\alpha)}
,\label{j}\end{equation}

\noindent which, at present time, i.e. $z=0$, it becomes

\begin{equation}
 j(0;\nu,\alpha) = 1 - \Omega_{k}^{(0)} + \frac{9 \nu}{2}
 \left( 1 - \frac{\nu}{\alpha} \right) \Omega_{gcg}^{(0)}.
 \label{j0}
\end{equation}

\noindent In getting this latter expression we have made use of the
constraint $\Omega_{gcg}^{(0)} + \Omega_{cdm}^{(0)} +
\Omega_{k}^{(0)} = 1$.

This parameter contains information regarding the sound speed of
the dark matter component \cite{Ch98}. Also, the use of the jerk
formalism infuses the kinematical analysis with a feature in that
all $\Lambda$CDM models are represented by a single value of the
jerk parameter $j = 1$. Therefore, the jerk formalism  enables us
to constrain and facilitates simple tests for departures from the
$\Lambda$CDM model in the kinematical manner \cite{R07}. In this
reference (~\refcite{R07} and references therein) it is
reported the following values for the jerk parameter: from the
type Ia supernovae (SNIa) data of the Supernova Legacy Survey
project gives $j = 1.32^{+1.37}_{-1.21}$, the X-ray galaxy cluster
distance measurements gives $j = 0.51^{+2.55}_{-2.00}$, the gold
SNIa sample data yields a larger value $j = 2.75^{+1.22}_{-1.10}
$, and the combination of all these three data set gives $j =
2.16^{+0.81}_{-0.75}$.

\begin{figure}[pb]
\centerline{\psfig{file=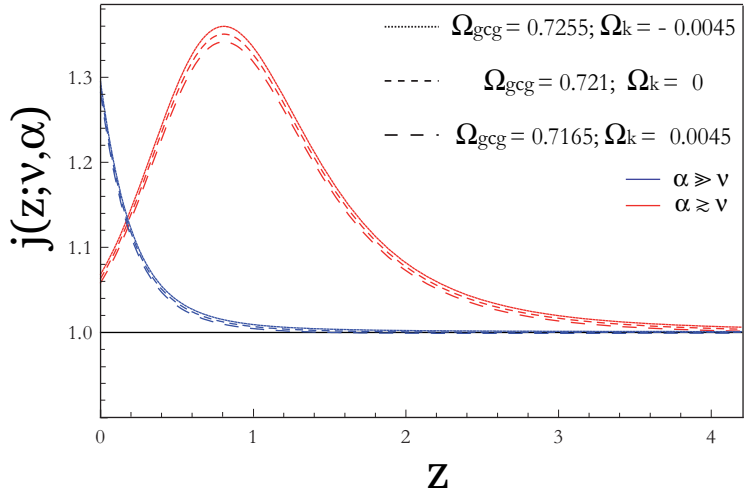,width=10cm}}
\vspace*{8pt}
  \caption{This plot presents the jerk, $j$, parameter as a
function of the redshifts, $z$. Here, we have taken the following
set of parameters:($\Omega_{gcg}^{(0)}=0.7255$ and
$\Omega_{k}^{(0)}=-0.0045$), ($\Omega_{gcg}^{(0)}=0.721$ and
$\Omega_{k}^{(0)}=0$) and ($\Omega_{gcg}^{(0)}=0.7165$ and
$\Omega_{k}^{(0)}=0.0045$). These three set of values have being
plotted for the two cases $\nu=0.1 \ll \alpha =0.9$ and $\nu =
0.88 \lesssim \alpha =0.9$.}
 \label{Fig07}
\end{figure}

In Fig. \ref{Fig07} we have plotted the jerk, $j$, parameter as a
function of the redshifts, $z$, for the set of parameters
($\Omega_{gcg}^{(0)}=0.7255$ and $\Omega_{k}^{(0)}=-0.0045$),
($\Omega_{gcg}^{(0)}=0.721$ and $\Omega_{k}^{(0)}=0$) and
($\Omega_{gcg}^{(0)}=0.7165$ and $\Omega_{k}^{(0)}=0.0045$). These
three set of parameters have being plotted for the two cases
$\nu=0.1 \ll \alpha =0.9.$ and $\nu = 0.88 \lesssim \alpha =0.9$.
Note that for the latter case the jerk function present a maximum
which is not present in the other case, when $\nu \ll \alpha $.
Note also that for $z \longrightarrow \infty$ the jerk parameter
goes to the value corresponding to the $\Lambda$CDM case. Also, we
do not observe much differences for the different type of
geometries, since the curves are very similar.

%$\Omega_{cdm}^{(0)}=.....$,

With respect to the snap parameter $s$ we have that this parameter
becomes given by

\begin{eqnarray} s(z;\nu,\alpha)&=&15 q^{3}(z;\nu,\alpha)
+ 9 q^{2}(z;\nu,\alpha) \nonumber \\& &- 10
q(z;\nu,\alpha) j(z;\nu,\alpha) - 3 j(z;\nu,\alpha) \nonumber \\&
& \hspace{2cm}-
\frac{(1+z)^{3}E'''(z;\nu,\alpha)}{E(z;\nu,\alpha)}
.\label{s}\end{eqnarray}

For a $\Lambda$CDM-universe the present expression for the snap
parameter becomes
$$
s_0=1-\frac{9}{2}\Omega_{cdm},
$$
and in our case it becomes at present, i.e. $z=0$,
\begin{eqnarray}
 s(0;\nu,\alpha)& =& \frac{9 \nu \Omega_{gcg}^{(0)}}{4 \alpha^{2}}
 \left[ 6 \alpha^{3} + \alpha^{2}(1- 18 \nu) + 3 \nu^{2}(2 -
 \Omega_{gcg}^{(0)})\right. \nonumber \\
 &+& \left. \alpha(2 + \nu (3 \Omega_{gcg}^{(0)} - 5 + 12 \nu)) \right] -
 \frac{7}{2}\\
 &+& \frac{\Omega_{k}^{(0)}}{4} \left[ 16 + 9 \nu \Omega_{gcg}^{(0)} - 2
 \Omega_{k}^{(0)} - 3\frac{\nu}{\alpha} (2 + 3\nu) \Omega_{gcg}^{(0)}\right]
 \nonumber.
\label{s0}
\end{eqnarray}
Here, we have used the constraint $\Omega_{gcg}^{(0)} +
\Omega_{cdm}^{(0)} + \Omega_{k}^{(0)} = 1$ also.

 In Ref.~\refcite{C08} was reported that the
actual value of the snap parameter, $s_0$, gets the value $s_0 =
3.39 \pm 17.13$ for the fit by using the LZ relation \cite{L05}
and the value $s_0 = 8.32 \pm 12.16$ for the fit by taking the GGL
one \cite{G04}.

\begin{figure}
\centerline{\psfig{file=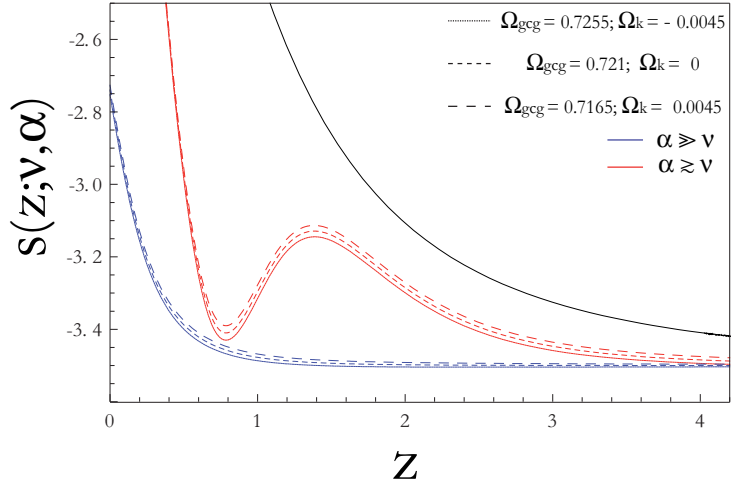,width=10cm}}
\vspace*{8pt}
  \caption{This plot presents the snap, $s$, parameter as a
function of the redshifts, $z$. Here, as before, we have taken the
following set of parameters:($\Omega_{gcg}^{(0)}=0.7255$ and
$\Omega_{k}^{(0)}=-0.0045$), ($\Omega_{gcg}^{(0)}=0.721$ and
$\Omega_{k}^{(0)}=0$) and ($\Omega_{gcg}^{(0)}=0.7165$ and
$\Omega_{k}^{(0)}=0.0045$). These three set of values have being
plotted for the two cases $\nu=0.1 \ll \alpha =0.9$ and $\nu =
0.88 \lesssim \alpha =0.9$.}
 \label{Fig08}
\end{figure}

\section{The first  Doppler peak of the CMB spectrum and
the shift parameter $R$}

In this section, we are going to describe the position of the
first Doppler peak ($l_{LS}^{gcg}$) for the  model studied in the
previous section. The scales that are important in determining the
shape of the CMB anisotropy spectrum are the sound horizon $d_s$
at the time of recombination, and the previously introduced
angular diameter distance $d_{A}^{LS}$ to the last scattering
surface. The former defines the physical scales for the Doppler
peak structure that depends on the physical matter density
($\Omega_{cdm}^{(0)}$), but not on the value of the GCG matter
density ($\Omega_{gcg}^{(0)}$ ) or spatial curvature
($\Omega_{k}^{(0)}$), since these are dynamically negligible at
the time of recombination \cite{E99}. The latter depends
practically on all of the parameters and is given by

\begin{equation}
d_{A}^{LS}= \frac{1}{H_{0}(1+z_{LS})} y(z_{LS}; \nu, \alpha)
\label{da}
\end{equation}

where $y(z_{LS}; \nu, \alpha)$ becomes given by
(see Eq. \ref{mo2})

\begin{equation}
y(z_{LS}; \nu, \alpha)= \frac{1}
{\sqrt{\left|\Omega_{k}^{(0)}\right|}} \,S_{k} \left\{
\sqrt{\left|\Omega_{k}^{(0)}\right|} \int_0^{z_{LS}} \frac{d z'}{
E(z';\nu,\alpha)} \right\}. \label{mo3}
\end{equation}

We may write for the localization of the first Doppler peak

\begin{equation}
l_{LS} \propto \frac{d_{A}^{LS}}{d_s} \label{ls}
\end{equation}

where the constant of proportionality depends on both the shape of the
primordial power spectrum and the Doppler peak number \cite{H96}.
Since we are going to keep the $\Omega_{cdm}^{(0)}$ parameter
fixed, we shall take $l_{LS}\approx d_A^{LS}$, up to a factor that
depends on $\Omega_{cdm}^{(0)}$ and $z_{LS}$ only

By using that
$\Omega_{k}^{(0)}=1-\Omega_{cdm}^{(0)}-\Omega_{gcg}^{(0)}$ and
following ref. ~\refcite{W00} and ref.~\refcite{d03} we can write for the
position of the first Doppler peak ($l_{LS}^{gcg}$)

\begin{equation}
l_{LS}^{gcg}\sim \Omega_{T}^{-\eta},\label{cmb1}
\end{equation}

\noindent where $\Omega_T =
\Omega_{k}^{(0)}+\Omega_{cdm}^{(0)}+\Omega_{gcg}^{(0)}$ and

\begin{equation}
\eta =
\frac{1}{6}I_{1}^{2}-\frac{1}{2}\frac{I_{2}}{I_{1}},\label{cmb2}
\end{equation}

\noindent with

\begin{equation}
I_{1}=\int_{0}^{1}\frac{dx}{\sqrt{(1-\Omega^{(0)}_{cdm})
x^{4}f_{gcg}(1/x-1;\nu,\alpha)+\Omega^{(0)}_{cdm}
x}},\label{cmb3}
\end{equation}

\noindent and

\begin{equation}
I_{2}=\int_{0}^{1}\frac{x^{4}f_{gcg}(1/x-1;\nu,\alpha)
dx}{\sqrt{\left[(1-\Omega^{(0)}_{cdm})x^{4}f_{gcg}(1/x-1;\nu,\alpha)+\Omega^{(0)}_{cdm}
x\right]^{3}}}.\label{cmb4}
\end{equation}

\noindent where $x=1/(1+z)$.

Note that the model $\Lambda CDM$ it is obtained  when
$f_{gcg}=1$, which corresponds to take the values $\alpha = \nu=0$
\cite{d03}.

\begin{figure}
\centerline{\psfig{file=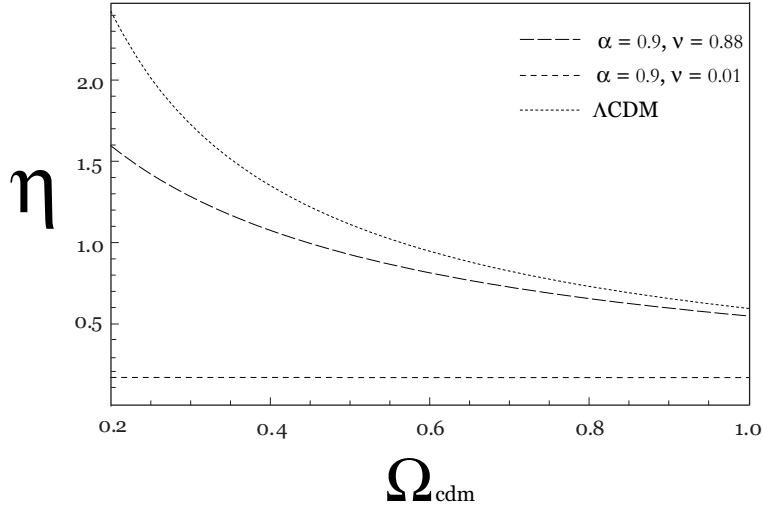,width=10cm}}
\vspace*{8pt}
\caption{This graph shows the parameter $\eta$ as a
function of the $\Omega_{cdm}^{(0)}$ parameter. We have considered
two different set of values for the gcg parameters: the set
($\alpha=0.9; \nu=0.88$) and the set ($\alpha=0.9; \nu=0.01$).
Here, we have included the $\Lambda$CDM case.}
  \label{fig:fig.eta}
\end{figure}

In FIG.\ref{fig:fig.eta} we show the parameter $\eta$ as a
function of the $\Omega_{cdm}^{(0)}$ parameter. Here, we have
taken two different set of values for the gcg parameters,
$\alpha=0.9; \nu=0.88$ and $\alpha=0.9; \nu=0.01$. In order to
make a comparison we have included in this plot  the $\Lambda$CDM
model.

One important parameter that describes the dependence of the first
Doppler peak position on the different parameters that
characterize any model is the shift parameter $R$. More specific,
it gives the position of the first Doppler peak with respect to
its location in a flat reference model with $\Omega_{cdm}^{(0)}=1$
\cite{B97,Tr04}. This becomes

\begin{equation}
R(\Omega_{cdm}^{(0)},\Omega_{gcg}^{(0)};\nu,\alpha) =\sqrt{ \frac{\Omega_{cdm}^{(0)}}{|\Omega_{k}^{(0)}|}}
S_{k}\left[\sqrt{|\Omega_{k}^{(0)}|}\int^{1}_{0}\frac{dx}{x^{2}
 E(x;\nu, \alpha)}\right], \nonumber \label{r2}
\end{equation}

\noindent where
$\Omega_{k}^{(0)}=1-(\Omega_{cdm}^{(0)}+\Omega_{gcg}^{(0)})$. Note
that the initial point is common for the same value of the
parameter with different curvature, and the final point is common
for the same curvature with different value of the parameters.
Note also that if we choose $\Omega_{k}^{(0)}=0$ and $\alpha=\nu =
0$ ($f_{gcg}(z;0,0)\rightarrow 1$) the $\Lambda$CDM case is
recuperated.

\begin{figure}
\centerline{\psfig{file=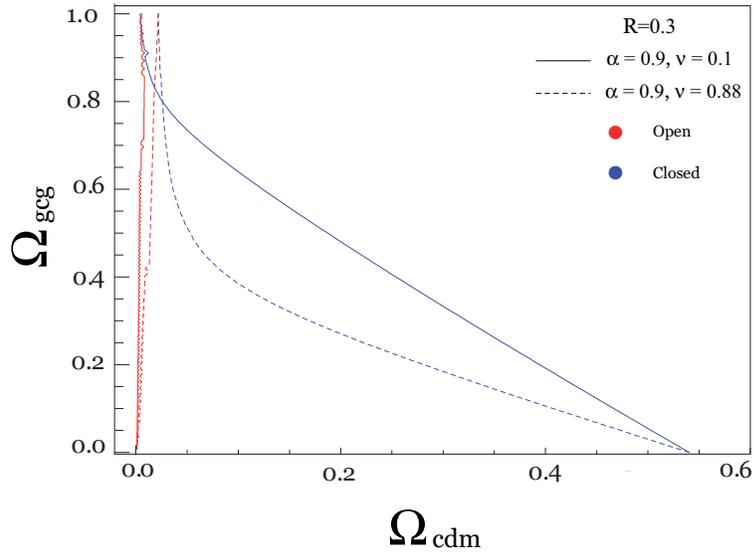,width=10cm}}
\vspace*{8pt}
\caption[]{Contour Plot in the $\Omega_{gcg}^{(0)}-
\Omega_{cdm}^{(0)}$ plane with $R = 0.3$ for
 two set of values for the parameters $\nu$ and
$\alpha$, i.e. $\nu = 0.1$ and $\nu = 0.88$ for $\alpha = 0.9$.
Here, we have considered positive and negative
curvature.}
\label{fig:fig.as}
\end{figure}

\section{Conclusions}
In this paper we have described and study a cosmological model in
which, apart from the usual cold dark matter component, we have
included a GCG associated to the dark energy component.  In this
kind of model we have described the properties of the GCG. The
characterization of the GCG comes from the determination of the
GCG parameters, $\nu$ and $\alpha$ related to the velocity of
sound of the fluid and the power appearing in the EOS of the GCG,
respectively. By taking into account some observational
astronomical data, such that the Hubble parameter, the
$y$-parameter, the angular size and the luminosity distance we
were able to restrict these parameters. All of them agree with the
condition $\nu\lesssim \alpha$. We have also described the
deceleration, the jerk and the snap parameters for our model. We
expect that with an appropriate data of these parameters will be
possible to restrict the parameters of the GCG fluid.

As an applicability of the GCG model described above, we have
determined the position of the first Doppler peak together with
the shift parameter R. These cases were compared with that
corresponding the $\Lambda$CDM model.

We may conclude that, as far as we are concerned with the observed
acceleration detected in the universe and the location of the
first Doppler peak, we will be able to utilize a GCG model to
describe the Universe we live in.

\end{document}